\documentclass[
preprint,
 amsmath,amssymb,
 aps,
]{revtex4-1}

\usepackage{graphicx}
\usepackage{dcolumn}
\usepackage{bm}
\usepackage{cmap}
\usepackage{epstopdf}
\usepackage{color}
\usepackage{verbatim}

\begin{document}

\preprint{APS/123-QED}

\title{The iridium double perovskite Sr$_2$YIrO$_6$ revisited: A combined structural and specific heat study}

\author{L.T. Corredor$^{1}$}
\thanks{these authors contributed equally to this work.}
\author{G. Aslan-Cansever$^{1*}$}
\author{M. Sturza$^{1*}$}
\author{Kaustuv Manna$^{1}$}
\author{A. Maljuk$^1$}
\author{S. Gass$^1$}
\author{A. Zimmermann$^1$}
\author{T. Dey$^1$}
\author{C. G. F. Blum$^1$}
\author{M. Geyer$^1$}
\author{A. U. B. Wolter$^1$}
\author{S. Wurmehl$^{1,2}$}
\author{B. B\"{u}chner$^{1,2}$}
\affiliation{$^1$Leibniz Institute for Solid State and Materials Research IFW, Institute for Solid
State Research, 01069 Dresden, Germany} \affiliation{$^2$Institute for solid state physics, Technische
Universit\"{a}t Dresden, D-01062 Dresden, Germany}

\date{}

\begin{abstract}

Recently, the iridate double perovskite Sr$_2$YIrO$_6$ has attracted considerable attention due to the
report of unexpected magnetism in this Ir$^{5+}$ (5d$^4$) material, in which according to the
J$_{eff}$ model, a non-magnetic ground state is expected. However, in recent works on polycrystalline
samples of the series Ba$_{2-x}$Sr$_x$YIrO$_6$ no indication of magnetic transitions have been found.
We present a structural, magnetic and thermodynamic characterization of Sr$_2$YIrO$_6$ single
crystals, with emphasis on the temperature and magnetic field dependence of the specific heat. Here,
we demonstrate the clue role of single crystal X-ray diffraction on the structural characterization of
the Sr$_2$YIrO$_6$ double perovskite crystals by reporting the detection of a $\sqrt{2}a \times
\sqrt{2}a \times 1c$ supercell, where $a$, $b$ and $c$ are the unit cell dimensions of the reported
monoclinic subcell. In agreement with the expected non-magnetic ground state of Ir$^{5+}$ (5d$^4$) in
Sr$_2$YIrO$_6$, no magnetic transition is observed down to 430~mK. Moreover, our results suggest that
the low temperature anomaly observed in the specific heat is not related to the onset of long-range
magnetic order. Instead, it is identified as a Schottky anomaly caused by paramagnetic impurities
present in the sample, of the order of $n \sim 0.5(2)$ \%. These impurities lead to non-negligible
spin correlations, which nonetheless, are not associated with long-range magnetic ordering.

\end{abstract}


\maketitle

\section{\label{intro}Introduction}

The effect of spin-orbit coupling (SOC) on strongly correlated
electron systems has demonstrated to be the origin of novel phases
\cite{Rau,Witczak}, like spin liquids \cite{Nag}, unconventional
magnetism \cite{Chikara,Ge}, and topological phases \cite{Chen2},
among others. Also, it has been demonstrated that for an
intermediate strength regime, SOC can work together with Coulomb
interactions to enhance the electron correlations, leading to
spin-orbit driven Mott insulators \cite{Jackeli,Pesin}. The key
point is that small changes in these interactions can lead to an
enormous variety in the electronic and magnetic behaviors.
Specifically in $5d$ transition metals like Ir the SOC is
significant, and comparable to the atomic interactions like
crystal electric field $\Delta$ and the on-site Coulomb
interaction $U$. Examples for this scenario are the pyrochlore
iridates, which show novel phases like Weyl semimetals
\cite{Wan,Witczak2,Burkov}, topological Mott insulators
\cite{Witczak3,Kargarian}, and axion insulators with unusual
electromagnetic responses \cite{Wan2,Go}.

The most common oxidation states of iridium are 3+ and 4+. On the
other hand oxidation states of 5+ and 6+ are rare and sometimes
poorly characterized \cite{Greenwood}. The double perovskite
structure R$_2$MM$^{\prime}$O$_6$ may host iridium on the M and/or
M$^{\prime}$ position; Ir will be in octahedral coordination in
this structure type. Hence, the octahedral crystal electric field
splits the $5d$ levels in a $t_{2g}$ triplet and a $e_g$ doublet.
Then, for Ir$^{4+}$ (5d$^5$ electronic configuration) the large
SOC acts on the $t_{2g}$ levels, splitting them into an effective
$j = 1/2$ doublet and an effective $j = 3/2$ quartet
\cite{Chen,Kim}.

According to the strong spin-orbit coupling J$_{eff}$ model, a non-magnetic $j= 0$ ground state is
expected for double perovskites  with Ir$^{5+}$ (5d$^4$) as the formal oxidation state. That is the
case for the monoclinic Sr$_2$YIrO$_6$, synthesized for the first time by Wakeshima et al.
\cite{Wakeshima} in 1999 with the aim of studying the magnetic properties of $5d$ transition metal
oxides in which the metal ions are in an unusual oxidation state. As expected, no transition into a
long-range ordered state was observed. However, recently a report by Cao et al. on this system
\cite{Cao} reports that a strong non-cubic crystal field (due to the flattening of the IrO$_6$
octahedra) together with an ``intermediate-strength'' spin-orbit coupling, leads to a different
ground-state configuration, i.e. an antiferromagnetic ground state with\linebreak T$_N$ = 1.3 K. The
emergence of magnetic ordering was evidenced in magnetization studies at 7 T as well as in specific
heat measurements in different applied magnetic fields. For the latter characterization, a small
anomaly in the low temperature region was interpreted as a signature of long-range magnetic ordering.

Recent first-principles calculations \cite{Bhowal} on both Sr$_2$YIrO$_6$ and Ba$_2$YIrO$_6$ compounds
gave another explanation for this behavior. In Ref \cite{Bhowal} the authors argue that the breakdown
of the J = 0 state in Sr$_2$YIrO$_6$ would be due to a band structure effect rather than the non-cubic
crystal field effect. However, a posterior report by Pajskr et al. \cite{Pajskr} combining several
numerical and semianalytical methods to study the band structure of both compounds contradicts the
conclusions of Refs. \cite{Cao} and \cite{Bhowal}. They found that monoclinic Sr$_2$YIrO$_6$ and cubic
Ba$_2$YIrO$_6$ are quite similar, and that both exhibit no tendency towards a magnetic phase
transition at low temperatures.

More recent works on polycrystalline samples of the series Ba$_{2-x}$Sr$_x$YIrO$_6$
\cite{Ranjbar,Phelan} did not observe any signature for long-range magnetic order in this material in
their magnetic susceptibility characterization, in strong contrast with Cao et al \cite{Cao}. On the
contrary, these reports reinforce the notion of a non-magnetic ground state dominated by strong SOC.
In both reports \cite{Ranjbar,Phelan} the magnetic characteristics of the system do not change
significantly across the series, and the effective magnetic moment per Ir is much less than the value
expected for a S = 1 system as reported by Cao et al. ~\cite{Cao}, demonstrating a strongly SOC
dominated ground state.

In this work, in order to settle the dispute described above, we present a combined structural,
magnetic and thermodynamic characterization of Sr$_2$YIrO$_6$ single crystals. Structure and
composition of the crystals were thoroughly characterized by single crystal X-ray diffraction (XRD),
synchroton powder XRD, energy dispersive X-ray spectroscopy (EDX) and scanning electron microscopy
(SEM). The single crystal XRD data show evidence of a cubic supercell. We particularly analyze the
specific heat of this material, in order to study the anomaly reported in \cite{Cao}. In our studies
no long-range magnetic order was found even in fields up to 9 T. The magnetic contribution to the
specific heat was extracted, finding a Schottky anomaly due to a small amount of magnetic impurities.
Further analysis suggests non-negligible spin correlations, which nonetheless, are not associated with
long-range magnetic ordering.
\newpage

\section{\label{sec:exp} Experimental details}

\subsection{Crystal growth}

Single crystals of Sr$_2$YIrO$_6$ were flux-grown using SrCO$_3$ (Alfa Aesar, 99.994\%), Y$_2$O$_3$
(Alfa Aesar, 99.99\%) and IrO$_2$ (Alpha Aesar, 99.99\%) as the precursor materials. Anhydrous
SrCl$_2$ (Alpha Aesar, 99.5\%) was used as the flux. The stoichiometric amount of starting materials
and the flux were mixed using a nutrient to solvent weight ratio 1:13. All constituents were put in a
platinum crucible (50 cubic ml) with a platinum lid. The crucible edge and lid were squeezed to
semi-seal the crucible assembly. The set-up was heated to different high temperatures
(1200$^{\circ}$C, 1250$^{\circ}$C and 1300$^{\circ}$C), held for 24 h and then cooled to
900$^{\circ}$C with different cooling rates (0.5$^{\circ}$C/h, 1$^{\circ}$C/h and 2$^{\circ}$C/h). The
furnace was switched off afterwards and cooled to room temperature.

The crystals were found at the bottom of the crucible and were separated by dissolving the flux in
water. After that, in order to gain a better understanding of the growth, the Sr$_2$YIrO$_6$ crystals
were examined by optical microscopy. As depicted in Figure S1 \cite{SM}, crystals have different
morphologies: cubic-like or irregular polyhedral. The cubic ones correspond to the pure Sr$_2$YIrO$_6$
phase and the irregular ones seem to be cubic crystals inter-grown with each other due to additional
nucleation sites. All our attempts to avoid the co-crystallization of irregular shaped crystals by
varying the hold temperatures, cooling rates and material to flux ratio have been failed. For the
following studies, we used the samples grown by cooling down from 1200$^{\circ}$C to 900$^{\circ}$C
with a rate of 2$^{\circ}$C/h and material to flux ratio of 1:13.

\subsection{Characterization}

A bunch of single crystals of Sr$_2$YIrO$_6$ with different morphologies were selected and mounted on
the tip of a glass fiber for X-ray Diffraction (XRD). Intensity data were collected at room
temperature using $\omega$ scans on a STOE imaging plate diffraction system (IPDS-II) with
graphite-monochromatized Mo K$\alpha$ radiation ($\lambda$ = 0.71069 \AA) operating at 50 kV and 40 mA
using a 34 cm diameter imaging plate. Individual frames were collected with a 4 min exposure time and
a 0.8$^{\circ} \omega$ rotation. X-AREA, X-RED, and X-SHAPE software packages \cite{soft1} were used
for data collection, integration, and analytical absorption corrections, respectively. SHELXL
\cite{soft2} and JANA2006 \cite{soft3} software packages were used to solve and refine the structure.

High-resolution synchrotron powder diffraction data were collected
at the beamline P02.1 at the storage ring PETRA III (DESY,
Hamburg, Germany) using an average wavelength of 0.2067
\AA\,($\sim$ 60 keV), with a relative energy bandwidths
$\Delta$E/E of the order of 10$^{-4}$. The repeatability of the
detector translations along the beam was tested in a series of
exposures of a LaB$_6$ standard (NIST 660a) filled into a 0.8
mm-diameter capillary. During this measurement, the detector was
moved forward and backward (with and without backlash) over the
entire travel range in steps of 125 mm \cite{Dippel}. The
synchrotron powder XRD pattern was collected on crushed single
crystals of Sr$_2$YIrO$_6$ sealed in a 0.8 mm-diameter capillary.
Data were fitted by the Rietveld method \cite{Rietveld} using
Fullprof in the WinPlotR program package \cite{Roisnel}.

The microstructural and compositional analysis was performed using scanning electron microscopy (SEM,
Zeiss EVOMA15) along with an electron microprobe analyzer for semi-quantitative elemental analysis in
the energy dispersive x-ray (EDX) mode (X-MaxN20 detector from Oxford Instruments with a AZtecEnergy
Advanced acquisition and EDX analysis software).

The magnetization as a function of temperature (in the range 0.43 - 300 K) and magnetic field was
obtained for randomly oriented single crystals of Sr$_2$YIrO$_6$ ($\sim$ 92.5 mg) using a
Superconducting Quantum Interference Device (SQUID) magnetometer from Quantum Design, equipped with an
{\it iHelium3} option. A thorough background subtraction was performed for all the curves. Specific
heat measurements were performed on 29 single crystals\linebreak ($\sim$ 5.6 mg) between 0.4 K and 10
K using a heat-pulse relaxation method in a Physical Properties Measurement System (PPMS) from Quantum
Design. The heat capacity of the sample holder (addenda) was determined prior to the measurements for
the purpose of separating the heat capacity contribution of the sample from the total heat capacity.

\section{\label{sec:res} Results and discussion}

Sr$_2$YIrO$_6$ orders in the double perovskite type structure with the general formula
R$_2$MM$^{\prime}$O$_6$, wherein R denotes an alkaline-earth (or rare-earth) metal and M and
M$^{\prime}$ are d-block elements (or other metals). Double perovskites crystallize in a cubic,
tetragonal, or monoclinic symmetry with interpenetrating M and M$^{\prime}$ face-centered cubic (FCC)
sublattices. We started the analysis of Sr$_2$YIrO$_6$ single crystal XRD data using the reported
monoclinic structural model \cite{Cao}, with the space group P2$_1$/n and lattice
parameters:\linebreak $a$ = 5.7826(5) \AA, b = 5.7830(5) \AA, c = 8.1746(7) \AA, $\beta$ =
90.036(7)$^{\circ}$. The single crystal XRD refinement leads to the agreement factors of R$_{obs}$ =
4.17 \% (for I $>2\sigma$(I)) and\linebreak R$_{all}$ = 9.19 \%.

\begin{figure}[hbt]
\includegraphics[scale=0.5]{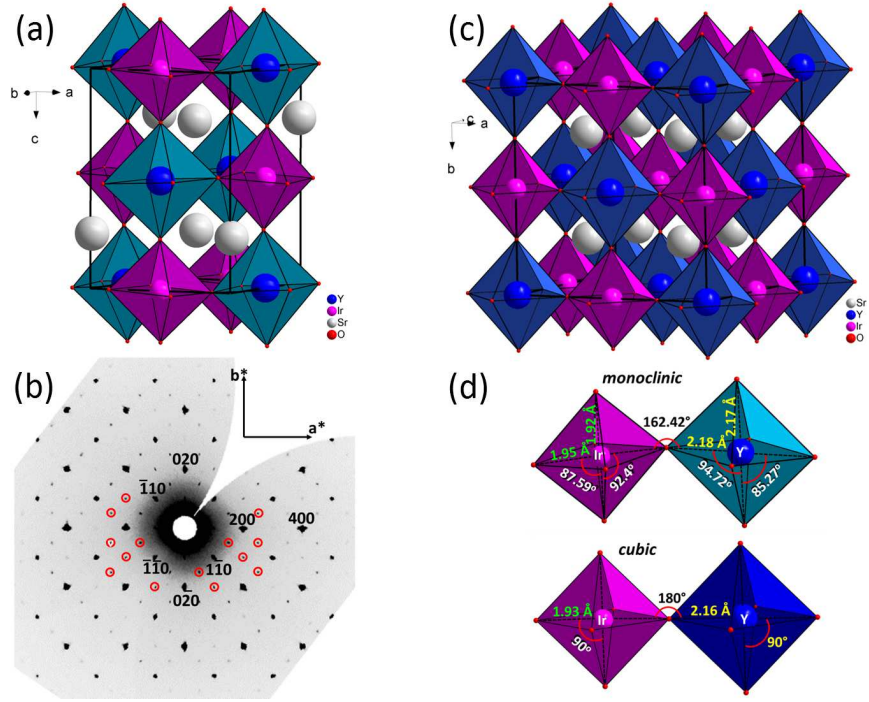}
\caption{\label{structure} (Color online) Structural details for
Sr$_2$YIrO$_6$ crystals: (a) Perspective view of the double
perovskite structure of the monoclinic subcell along the
crystallographic b axis. (b) Reconstructed precession photographs
for Sr$_2$YIrO$_6$ crystal in the $hk0$ layers show extra
reflections, marked with red circles, required to describe a
$\sqrt{2} \times \sqrt{2} \times 1$ supercell. (c) Perspective
view of the double perovskite structure of the $\sqrt{2} \times
\sqrt{2} \times 1$ cubic supercell along the crystallographic c
axis. (d) Comparison between bond lengths and bond angles of the
IrO$_6$ and YO$_6$ octahedra, monoclinic vs cubic.}
\end{figure}

After an anisotropic refinement of all atoms, a difference
electron density Fourier map calculated with phases based on the
final parameters shows maximum and minimum residual peaks of +2.03
and -2.58 e\AA$^{-3}$, respectively. A detailed examination of
reconstructed precession images of the $hk0$ layer (Figure
\ref{structure}(b)) shows strong diffraction maxima due to the
small subcell and additional weak peaks forming distinctive groups
of four reflections around the allowed peaks. All those extra
peaks can be indexed with the $\sqrt{2}a \times \sqrt{2}b \times
1c$ supercell, where $a$, $b$ and $c$ are the unit cell dimensions
of the monoclinic subcell. The integrated data with the $\sqrt{2}a
\times \sqrt{2}b \times 1c$ supercell suggests that Sr$_2$YIrO$_6$
crystallizes in the cubic space group $Fm\overline{3}m$ with
lattice parameter, $a$ = 8.1773(7) \AA.

The crystal structure of the supercell based on our refinement, shown in Figure \ref{structure}(c), is
built up of corner-sharing IrO$_6$ and YO$_6$ octahedra units, which feature an ordered rock-salt like
arrangement, and the Sr placed in between. Please note that the Ir$^{5+}$ ions in this structure build
up a FCC network, a general feature of the double perovskite type structure. The single crystal XRD
refinement by taking into account the supercell shows excellent agreement factors of R$_{obs}$ = 1.65
\% (for I $>2\sigma$(I)) and R$_{all}$ = 3.79 \% and less electron density difference after an
anisotropic refinement of all atoms, i.e. maximum and minimum residual peaks +1.06 and -0.98
e\AA$^{-3}$, respectively. The parameters for data collection and the details of the structure
refinement are given in Table \ref{table1}. Atomic coordinates, thermal displacement parameters
(U$_{eq}$) and occupancies of all atoms are given in Table \ref{table2}. Anisotropic displacement
parameters (ADPs) and selected bond lengths and angles are given in the supplemental material, Tables
S1 and S2 \cite{SM}.

\begin{table}[!htbp]
\caption{\label{table1}\footnotesize Crystal data and structure refinement for Sr$_2$YIrO$_6$ (a)
subcell and (b) supercell at 293 K.}
\begin{ruledtabular}
\begin{tabular}{ccc}
 & (a) subcell & (b) supercell\\\hline
Empirical formula & Sr$_2$YIrO$_6$ & Sr$_2$YIrO$_6$\\
Formula weight & 552.35 & 552.35\\
Temperature & 293(2) K & 293(2) K\\
Wavelength & 0.71069 \AA & 0.71069 \AA\\
Crystal system & monoclinic & cubic\\
Space group & P$2_1/n$ & $Fm\overline{3}m$\\
 & $a$ = 5.7826(5) \AA, $\alpha$ = 90$^{\circ}$ & $a$ = 8.1773(7) \AA, $\alpha$ = 90$^{\circ}$\\
Unit cell dimensions & $b$ = 5.7830(5) \AA, $\beta$ = 90.036(7)$^{\circ}$ & $b$= 8.1773(7) \AA, $\beta$ = 90$^{\circ}$\\
 & c = 8.1746(7) \AA, $\gamma = 90^{\circ}$ & c = 8.1773(7)  \AA, $\gamma = 90^{\circ}$\\
Volume & 273.36(4) \AA$^3$ & 546.80(14) \AA$^3$\\
Z & 2 & 4\\
Density (calculated) & 6.710 g/cm$^3$ & 6.710 g/cm$^3$\\
Absorption coefficient & 54.136 mm$^{-1}$ & 54.129 mm$^{-1}$\\
F(000) & 480 & 960\\
$\Theta$ range for data collection & 4.316 to 24.927$^{\circ}$ & 4.317 to 26.846$^{\circ}$\\
Crystal size & $0.042 \times 0.034 \times 0.03$3 mm$^3$ & $0.042 \times 0.034 \times 0.033$ mm$^3$\\
Reflections collected & 1659 & 1128\\
Independent reflections & 477[R$_{int}$ = 0.0337] & 52[R$_{int}$ = 0.0488]\\
Completeness to $\Theta = 24.97^{\circ}$ & 97.9 \% & 97.6 \%\\
Data / restraints / parameters & 477 / 0 / 50 & 52 / 0 / 8\\
Goodness-of-fit & 1.376 & 1.381\\
Final R indices [$>2\sigma(I)$] & R$_{obs}$ = 0.0368\footnotemark[1], wR$_{obs}$ = 0.0906\footnotemark[2] & R$_{obs}$ = 0.0135\footnotemark[1], wR$_{obs}$ = 0.0359\footnotemark[2]\\
R indices [all data] & R$_{all}$ = 0.0417\footnotemark[1], wR$_{all}$ = 0.0919\footnotemark[2] & R$_{all}$ = 0.0165\footnotemark[1], wR$_{all}$ = 0.0379\footnotemark[2]\\
\end{tabular}
\end{ruledtabular}
\footnotetext[1]{$R = \sum \mid\mid F_o\mid - \mid F_c\mid\mid/ \sum\mid F_o\mid$}
\footnotetext[2]{${wR = \{\sum[w(\mid F_o\mid^2 - \mid F_c\mid^2)^2]/\sum[w(\mid
F_o\mid^4)]\}^{1/2}}$}
\end{table}

\begin{table}[hbt]
\caption{\label{table2}\footnotesize Atomic coordinates ($\times
10^4$) and equivalent isotropic displacement parameters (\AA
$\times 10^3$) of Sr$_2$YIrO$_6$ (a) subcell and (b) supercell at
293(2) K with estimated standard deviations in parentheses.}
\vspace{0.5cm}
\begin{ruledtabular}
\begin{tabular}{ccccccc}
(a) Sr$_2$YIrO$_6$ subcell & Wyck & x & y & z & Occupancy & U$_{eq}^*$\\\hline
Y(1) & 2d & 0 & 5000 & 0 & 1 & 7(1)\\
Ir(1) & 2d & 5000 & 0 & 0 & 1 & 5(1)\\
Sr(1) & 4e & 5009(3) & 5148(3) & 2496(2) & 1 & 17(1)\\
O(1) & 4e & 2380(2) & 2100(3) & 20(3) & 1 & 59(7)\\
O(2) & 4e & 2980(3) & 7300(3) & -180(3) & 1 & 57(6)\\
O(3) & 4e & 4980(5) & -120(5) & 2340(2) & 1 & 84(8)\\\hline
(b) Sr$_2$YIrO$_6$ supercell & Wyck & x & y & z & Occupancy & U$_{eq}^*$\\\hline
Sr(1) & 8c & 2500 & 2500 & 2500 & 1 & 20(1)\\
Ir(1) & 4b & 5000 & 5000 & 5000 & 1 & 5(1)\\
Y(1) & 4a & 0 & 0 & 0 & 1 & 7(1)\\
O(1) & 24e & 5000 & 5000 & 7369(13) & 1 & 85(5)\\
\end{tabular}
\end{ruledtabular}
\end{table}

In the Sr$_2$YIrO$_6$ subcell, the monoclinic distortion is very
small, as presented in Figure \ref{structure}(d), with $\beta$
angles very close to 90$^{\circ}$ ($\beta = 90.036(7)^{\circ}$)
and unequal Ir-O bond lengths with a small difference of ~0.03
\AA, while $\angle$O-Ir-O deviates from 90$^{\circ}$ by
2.4$^{\circ}$, and the IrO$_6$ octahedra are rotated/tilted with
$\angle$Ir-O-Y of 162.42$^{\circ}$. For our Sr$_2$YIrO$_6$
supercell, the IrO$_6$ octahedra are completely regular with equal
Ir-O bond lengths, $\angle$O-Ir-O= 90$^{\circ}$ and $\angle$Ir-O-Y
= 180$^{\circ}$. Our experimental results on single crystal XRD
are in good agreement with the theoretical paper reported by
Pajskr et al. \cite{Pajskr}, which claims that the non-cubic
crystal field of Sr$_2$YIrO$_6$ is found to be rather weak and
that Sr$_2$YIrO$_6$ is quite similar in terms of structure and
low-energy properties to the cubic Ba$_2$YIrO$_6$ analog
\cite{Dey}.

The results of our Rietveld refinement, agreement factors, and refined lattice constants of
synchrotron powder XRD ($\lambda$ = 0.2067 \AA) patterns of the Sr$_2$YIrO$_6$ sample studied in this
work are shown in the supplemental material \cite{SM}. Figure S2 shows the experimental synchrotron
XRD data, the simulated and residual intensities as well as the corresponding Bragg positions using
the monoclinic structure model. A tiny trace of unreacted\linebreak Y$_2$O$_3$ ($\sim$ 1.06 \%) is
observed. The analysis of the structural data reveals that Sr$_2$YIrO$_6$ crystallizes in monoclinic
P2$_1$/n space group with lattice parameters: $a$ = 5.7870(7) \AA,\linebreak b = 5.7912(6) \AA, c =
8.1805(1) \AA, $\beta$ = 90.214(6)$^{\circ}$ and R$_f$ = 2.11, R$_{Bragg}$ = 2.01 and reduced $\chi^2$
= 2.32. The result of the Rietveld refinement of the synchrotron XRD data using the cubic structural
model discovered by single crystal XRD is depicted in Figure S3. Please note that both the monoclinic
and the cubic model describe the synchrotron powder XRD data reasonably well with a slightly better
fit by the monoclinic model (Bragg R-factor 2.11 for the monoclinic vs 3.46 for the cubic one).
However, this difference is not significant since the larger number of fit parameters for the
monoclinic model will yield better fit results as such. Hence, is very difficult to decide which
model, monoclinic or cubic, is correct using powder diffraction data only. We emphasize that single
crystal diffraction data is required to find the cubic $\sqrt{2}a \times \sqrt{2}b \times 1c$
supercell.

\begin{figure}[hbt]
\includegraphics[scale=0.17]{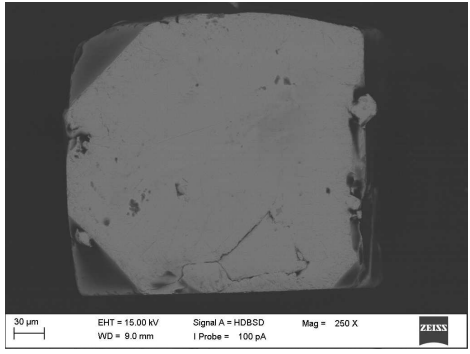}
\caption{\label{SEM} SEM image of a typical Sr$_2$YIrO$_6$ single
crystal in the backscattered electron (BSE) mode.}
\end{figure}

Figure \ref{SEM} illustrates an SEM image of the Sr$_2$YIrO$_6$
single crystal, collected in backscattered electron (BSE) mode.
EDX analysis was performed on various spots and on several
crystals. The analysis suggests that the crystals are chemically
homogeneous with a stoichiometry close to the target composition,
within the accuracy of the EDX method.
\newpage

\begin{figure}[htb]
\includegraphics[scale=0.4]{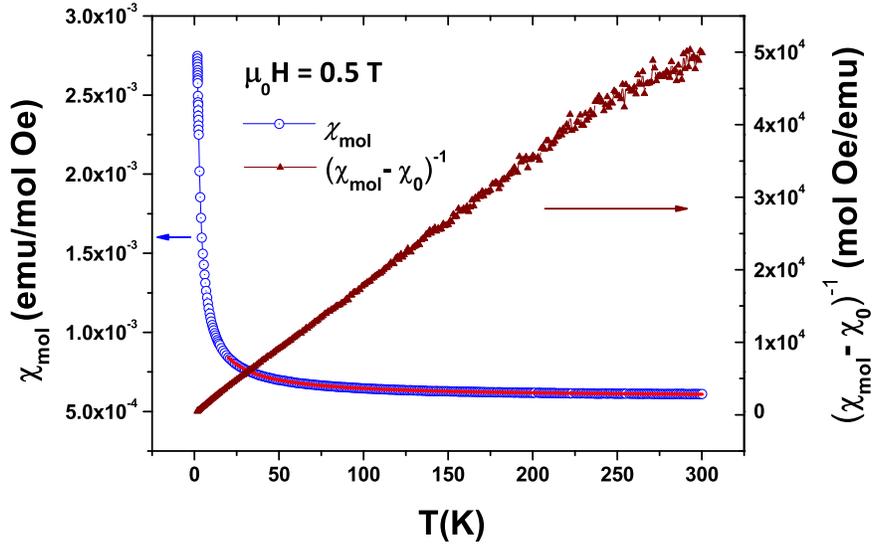}
\caption{\label{ChixT} (color online). The magnetic susceptibility of Sr$_2$YIrO$_6$ as a function of
temperature in an external magnetic field $\mu_0$H = 0.5 T, along with its fit according to a
Curie-Weiss (CW) law in the temperature range 20-300 K (left axis). On the right axis, the inverse
susceptibility (after subtracting $\chi_0$ obtained from the CW fit) is plotted.}
\end{figure}

The magnetic susceptibility in an external magnetic field of
$\mu_0$H = 0.5 T is shown in Fig. \ref{ChixT} in the temperature
range 1.8 -300 K, with no evidence of long-range magnetic order.
The data obeys the Curie-Weiss law $\chi(T) = \chi_0 +
C/(T-\Theta)$ in the temperature range 20 - 300 K, which gives a
temperature independent susceptibility contribution $\chi_0 = 5.90
\times 10^{-4}$ emu/mol Oe, an effective magnetic moment
$\mu_{eff} = 0.21 \mu_B$/Ir, and a Curie-Weiss
temperature\linebreak $\Theta = -2.8$ K.

Due to the $5d^4$ electronic configuration of Sr$_2$YIrO$_6$, a
Van Vleck contribution to the susceptibility is expected: $\chi_0
= \chi_{Dia} + \chi_{VV}$, where $\chi_{Dia}$ comes from the
diamagnetic contribution of the core levels and $\chi_{VV}$ is the
Van Vleck paramagnetic susceptibility. The diamagnetic
susceptibility can be obtained by adding the diamagnetic
contribution of all the individual ions, leading to $\chi_{Dia} =
-7.03 \times 10^{-5}$ emu/mol Oe, and finally $\chi_{VV} = 6.6
\times 10^{-4}$ emu/mol Oe, which is in agreement with similar
$5d^4$ compounds \cite{Dey,Bremholm}. It is worth to notice that
the obtained effective magnetic moment $\mu_{eff} = 0.21 \mu_B$/Ir
is much smaller than the value 2.38$\mu_B$/Ir expected for a
conventional S = 1 5d-electron system as mentioned by Cao et
al.~\cite{Cao}, but it is in close agreement with the value
$\mu_{eff} = 0.16 \mu_B$/Ir reported for polycrystalline
Sr$_2$YIrO$_6$ \cite{Ranjbar}. As will be shown below, the
paramagnetic signal can be attributed to a small percentage of
paramagnetic centers (less than 1\%), so it is likely that the
Curie-tail in the magnetic susceptibility stems from paramagnetic
impurities in the Sr$_2$YIrO$_6$ sample, only.

\begin{figure}[htb]
\includegraphics[scale=0.4]{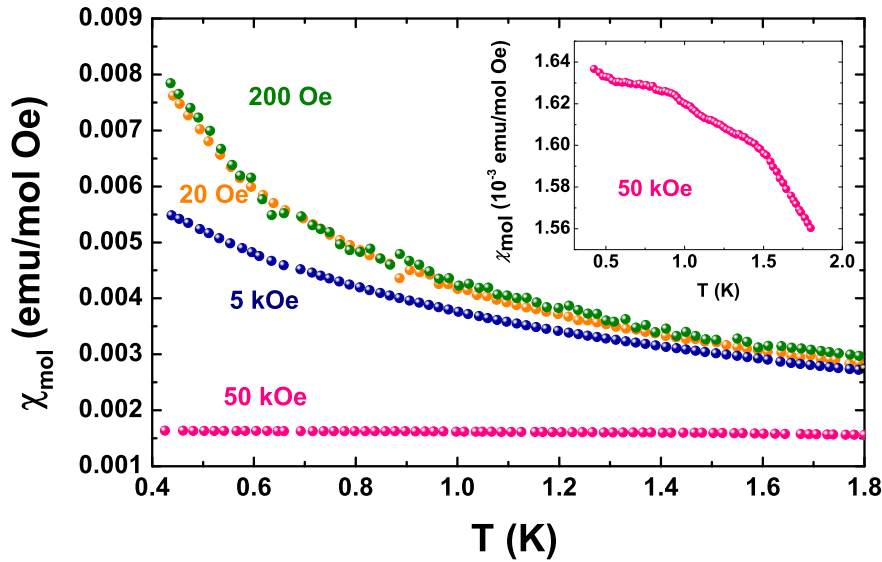}
\caption{\label{MxT} (color online). Temperature dependence of the
magnetic susceptibility in the low temperature region for
different magnetic fields (ZFC results). The inset shows a zoom of
the 5 T data down to 0.43~K.}
\end{figure}

In view of the antiferromagnetic transition at T$_N$ = 1.3 K in 7 T magnetic field which has been
observed in Ref.~\cite{Cao}, we performed further magnetization studies at low temperatures and at
high magnetic fields. From our experimental data no magnetic transition is detected down to 0.43~K,
even for applied magnetic fields up to 5~T (see Fig.\ref{MxT}) \cite{5T_note}.

\begin{figure}[hbt]
\includegraphics[scale=0.35]{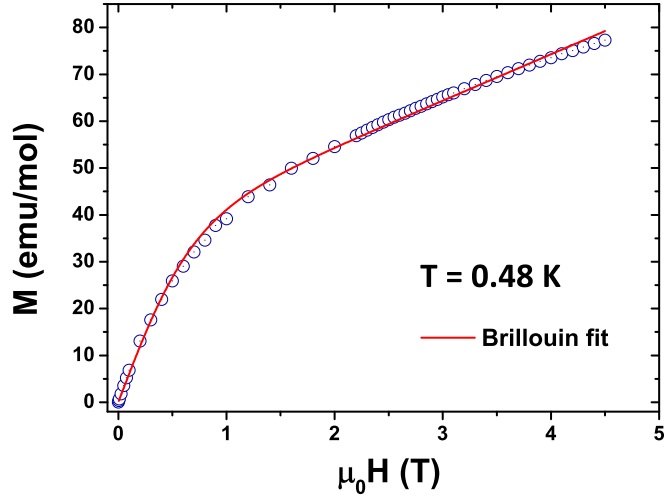}
\caption{\label{Brillouin} (color online). Isothermal magnetization curve of Sr$_2$YIrO$_6$ at T =
0.48 K. The red line shows the fit according to the Brillouin function (for details see text).}
\end{figure}

The magnetization of Sr$_2$YIrO$_6$ as a function of applied
magnetic field is shown in Fig. \ref{Brillouin} for T = 0.48~K. In
agreement with the temperature dependent measurements, the
magnetic behavior at lowest temperatures is predominantly
paramagnetic, and the curve can be fitted with the modified
Brillouin function:

\begin{equation}
M(H) = \chi_0 H + nN_Ag\mu_BJ\left\{
\frac{2J+1}{2J}\coth\Bigg(\frac{2J+1}{2J}\frac{g\mu_BJH}{k_BT}\Bigg)-\frac{1}{2J}\coth\Bigg(\frac{1}{2J}\frac{g\mu_BJH}{k_BT}\Bigg)\right\}.
\end{equation}

In this expression $n$ represents a scaling factor which gives the percentage of paramagnetic Ir in
the sample, $N_A$ is the Avogadro constant, $g$ the Land\'{e} factor, $\mu_B$ the Bohr magneton, $J$
the total angular momentum, and $k_B$ the Boltzmann constant. From the fit the parameters g = 2.1, and
$\chi_0 = 9.9 \times 10^{-4}$ emu/mol Oe are found, together with a low percentage of paramagnetic
centers ($J = 1/2$) of $n \sim 0.6$\%. These paramagnetic centers could be defects in the sample
stemming from oxygen vacancies, chemical disorder (Y/Ir site mixing), off-stoichiometry, etc. A
possible scenario could be the presence of Ir$^{4+}$ or Ir$^{6+}$ ions created by intermixing and/or
off-stoichiometry \cite{Dey,Kayser,g_note}.

The temperature independent susceptibility value $\chi_0$ is slightly larger than the value obtained
from the Curie-Weiss fitting Fig. \ref{ChixT}. This discrepancy could be due to non-negligible spin
correlations at low temperature which are not included in Eq.~\ref{Brillouin}, or to a
temperature-dependent contribution to the Van Vleck susceptibility. It is worth to notice that no
metamagnetic features arise in the field dependence of the magnetization in contrast to
Ref.~\cite{Cao}, where a sharp metamagnetic transition has been observed at the critical field H$_c
\approx$ 2.6 K at T = 0.5 K.

\begin{figure}[hbt]
\includegraphics[scale=0.35]{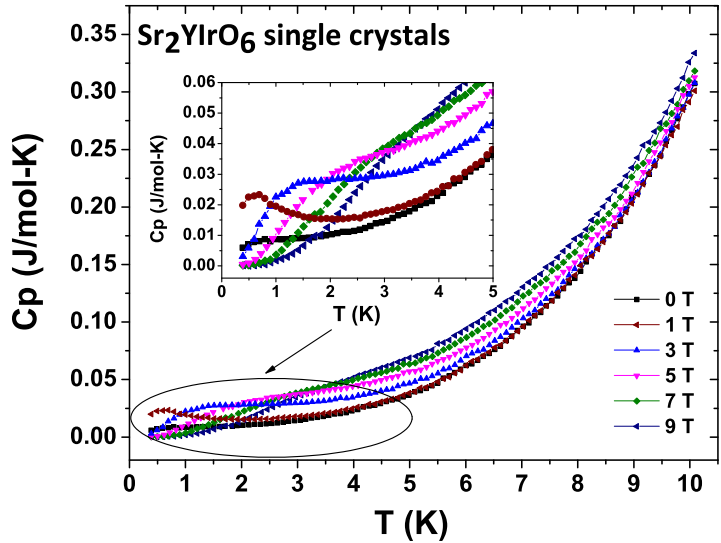}
\caption{\label{Cp_all} (color online). Temperature dependence of the specific heat of Sr$_2$YIrO$_6$
for different magnetic fields, shown in the same scale as \cite{Cao}. The inset shows a zoom into the
low-temperature region.}
\end{figure}

The low temperature specific heat in different magnetic fields is
shown in Fig. \ref{Cp_all}. A small anomaly is observed below T
$\sim$ 5 K in all the curves, including the zero-field data, and
which becomes more pronounced in applied magnetic fields of 1, 3,
and 5 T. This peak shifts to higher temperature and broadens with
increasing magnetic fields, which in principle suggests its
magnetic origin. However, its magnitude is small, and instead of
being a sharp, $\lambda$-like anomaly, pointing towards a
second-order phase transition, it resembles more a hump-like
feature. This makes it difficult to identify the anomaly as a real
magnetic phase transition.

In Ref.~\cite{Cao}, Cao et al. detect a similar anomaly for their Sr$_2$YIrO$_6$ single crystals, in
qualitatively and semi-quantitative agreement with the one observed here. In \cite{Cao}, the anomaly
in the specific heat was interpreted as the onset of long-range magnetic order, due to its proximity
to the ordering temperature T$_N$ = 1.3 K observed as a sharp peak in the magnetization as function of
temperature. In order to settle this open question, we performed additional specific heat measurements
on the non-magnetic cubic analog compound Ba$_2$YIrO$_6$ ~\cite{Dey}. Since we observe striking
similarities in our specific heat data for Sr$_2$YIrO$_6$ and Ba$_2$YIrO$_6$ \cite{SM}, i.e. a broad
anomaly below $T$ $\sim$ 5~K, it is rather improbable that the specific heat anomaly in Sr$_2$YIrO$_6$
marks a transition into a magnetically long-range ordered state.

\begin{figure}[h]
\includegraphics[scale=0.35]{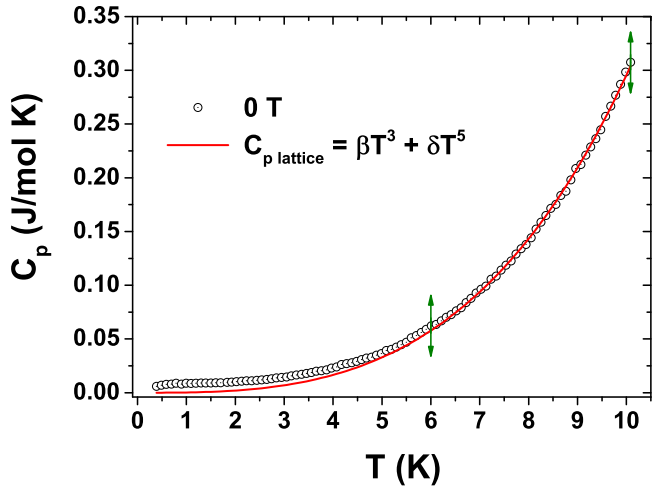}
\caption{\label{Lattice}(color online). Zero-field specific heat
of Sr$_2$YIrO$_6$. The red line represents the lattice specific
heat obtained using a power-law fitting in the range of 6 K $\leq$
T $\leq$ 10 K marked by arrows; for details see text.}
\end{figure}

To gain a deeper insight into the origin of this anomaly, the
lattice contribution was modelled for 6 K $\leq$ T $\leq$ 10 K
using the approximation $C_{p,lattice} = \beta T^3 + \delta T^5$
in order to disentangle possible magnetic contributions in
Sr$_2$YIrO$_6$ (Fig. \ref{Lattice}). The extracted magnetic
contribution shows a broad anomaly. From the fit parameters $\beta
= 2.52\times10^{-4}$ J/mol K$^4$ and $\delta = 4.36\times10^{-7}$
J/mol K$^6$, the Debye temperature $\Theta_D = 425.7$ K was
calculated. This value is comparable with those obtained for other
double-perovskite materials \cite{Du,Blum} and iridium-based
compounds \cite{Bremholm,Cai}.

\begin{figure}[h]
\includegraphics[scale=0.35]{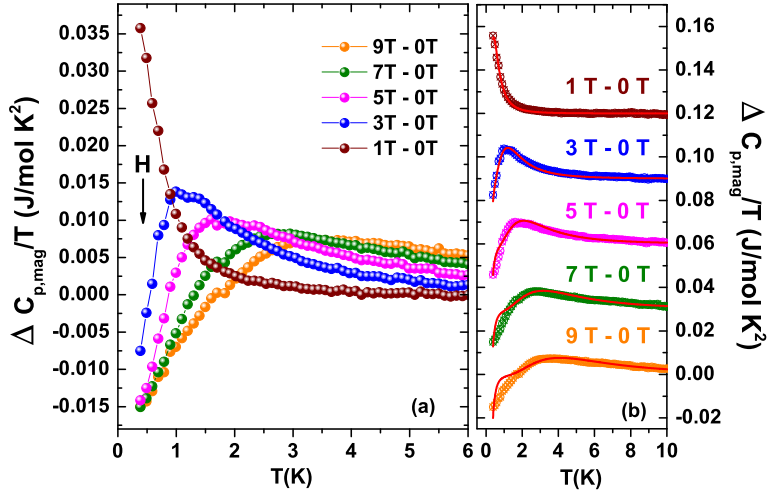}
\caption{\label{Schottky}(color online). (a) Temperature
dependence of the low temperature specific heat divided by
temperature for different magnetic fields, after subtraction of
the zero-field data. (b) Experimental data from (a) represented
with a constant offset of 0.03 J/mol K$^2$ for clarity. The red
lines represent the fits according to a two-level Schottky anomaly
(see text).}
\end{figure}

Considering that the fitting range was small and chosen to follow the best-fitting procedure in
combination with sufficient statistics, the extracted magnetic contribution to the specific heat has
only a semi-quantitative character, i.e. the assumption $\Delta C_{p,mag}$ = 0 for\linebreak $T$
$\geq$ 6~K has been applied. Thus, in order to model the magnetic contribution to $C_p$ in the
following a subtraction of the zero-field data from the corresponding field data was performed,
yielding the field-dependent magnetic specific heat. The subtracted data are shown in Fig.
\ref{Schottky}(a). The shape of the $\Delta C_{p,mag}/T$ curves resembles a two-level Schottky
anomaly, shifting to higher temperature with increasing magnetic field.

Consequently, the magnetic contribution shown in Fig.
\ref{Schottky}(a) was modelled as the subtraction of two Schottky
anomalies, i.e. $C_{m}(T,H) = C_{Sch}(T,H) - C_{Sch}(T,0)$.
Following this reasoning, the peaks were fitted in the range 0.4 K
$\leq$ T $\leq$ 6 K with the following expression:

\begin{equation}\label{Sch_eq}
C_{m}(T,H) = n\Bigg[\Bigg(\frac{\Delta(H)}{T}\Bigg)^2
\frac{e^{\Delta(H)/T}}{(1+e^{\Delta(H)/T})^2} -
\Bigg(\frac{\Delta_0}{T}\Bigg)^2
\frac{e^{\Delta_0/T}}{(1+e^{\Delta_0/T})^2}\Bigg],
\end{equation}
where $n$ is the concentration of paramagnetic impurities, $\Delta_0$ is the energy separation between
the two levels in zero magnetic field, and $\Delta(H) = g\mu_BH$ represents the Zeeman splitting in an
applied magnetic field $H$. The fits are shown in the Fig. \ref{Schottky}(b). According to the fit, a
value $\Delta_0$ = 0.25 K was found for the gap at zero field together with $n \approx 0.7$ \%,
indicating that the anomaly is only based on a small amount of paramagnetic impurities, instead of
being due to a real long-range magnetic ordering of all iridium atoms in Sr$_2$YIrO$_6$.

While the fits reasonably describe our data for H $\leq$ 5 T, a slightly different behavior is found
for H $\geq$ 7 T with a lacking peak at low temperatures in our experimental data. This suggests that
even when a Schottky anomaly can explain the observed peaks for H $\leq$ 5 T (low fields), for higher
fields the magnetic contribution is altered. Following the discussion about the Brillouin fitting in
Fig. \ref{Brillouin}, this could point towards field-induced changes, possibly due to non-negligible
spin correlations from correlated magnetic impurities which, nonetheless, are not associated with
long-range magnetic ordering, but with short-range correlations only. The possibility of magnetic
correlations due to impurities was already pointed out for polycrystalline Ba$_{2-x}$Sr$_x$YIrO$_6$
\cite{Ranjbar} and Ba$_2$YIrO$_6$ single crystals \cite{Dey}.

\begin{figure}[h]
\includegraphics[scale=0.35]{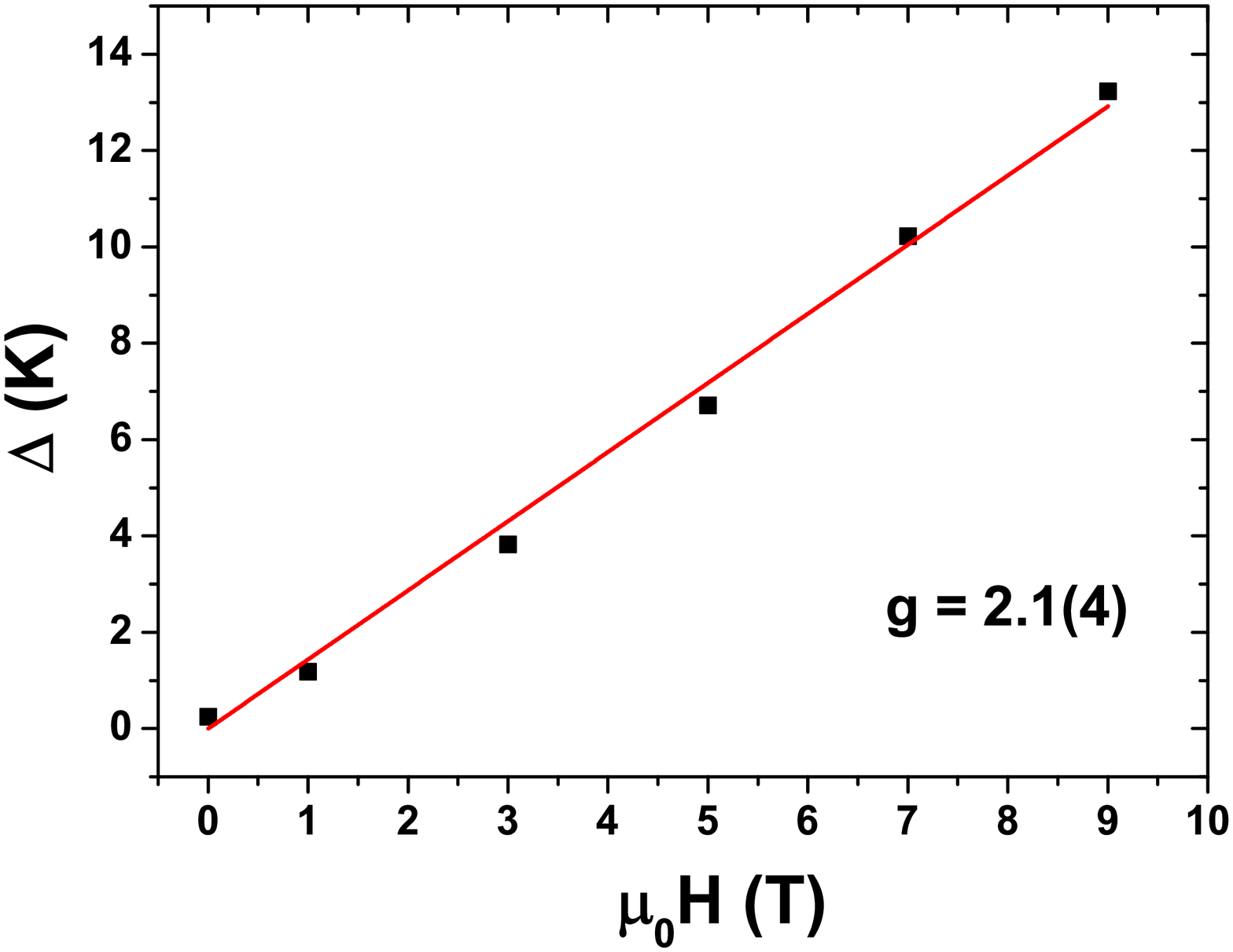}
\caption{\label{Gap}(color online). Magnetic field dependence of
the energy gap between the ground state doublet for paramagnetic
impurity spins in Sr$_2$YIrO$_6$. The red line and the $g$ value
are the fitting results using $\Delta = g\mu_BH$.}
\end{figure}

We emphasize the fact that both the paramagnetic susceptibility
and the Schottky two-level anomaly can be ascribed to the presence
of correlated impurity spins Ir$^{4+}$ or Ir$^{6+}$. With this in
mind, the Zeeman splitting between the two lowest energy levels
obtained from the former fits can be plotted as a function of
applied magnetic field (see Fig. \ref{Gap}). From $\Delta =
g\mu_BH$ the Land\'{e} $g$ factor was found to be $g$ = 2.1(4),
which is consistent with the $g$ factor obtained from the
Brillouin fit, $g$ = 2.1(0).

\begin{figure}[hbt]
\includegraphics[scale=0.35]{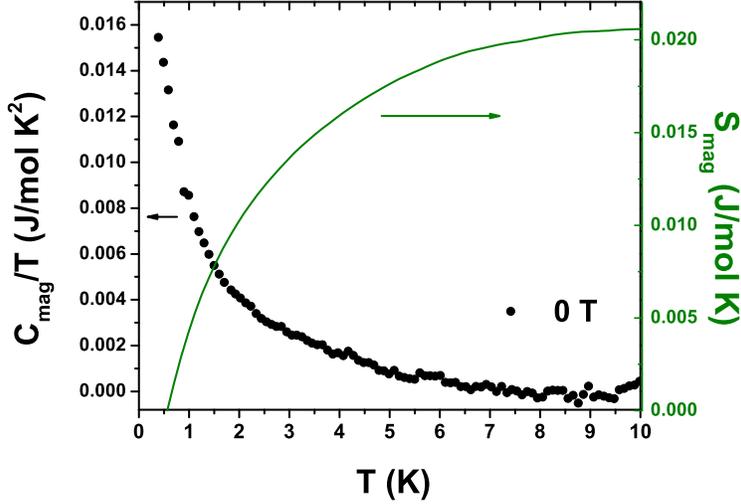}
\caption{\label{Entropy}(color online). Zero-field magnetic
specific heat plotted as C$_{p,mag}$/T vs T (left scale) together
with the magnetic entropy (right scale).}
\end{figure}

Finally, we can evaluate the magnetic entropy $S_{mag}$ for our
samples by integrating the zero-field specific heat data from Fig.
\ref{Lattice} according to the expression:

\begin{equation}
S_{mag} = \int_{0}^{T}  \! \frac{C_{mag}}{T} \, dT.
\end{equation}

The resulting curves for C$_{p,mag}$/T and S$_{mag}$(T) are shown in Fig. \ref{Entropy}. The small
value $S_{mag} = 0.021$ J/mol K is of the same order of magnitude as reported by Cao et al.
\cite{Cao}\linebreak ($\sim 0.01$ J/ mol K). This non-zero value again points to the presence of
paramagnetic impurities, which need to be taken into account. For a $S$ = 1/2 ground state of
Sr$_2$YIrO$_6$ as claimed in Ref.~\cite{Cao}, a total entropy $S_{mag} = R\ln(2J+1) = R\ln(2) = 5.76$
J/mol K is expected, which is evidently far from the experimental observations. Nevertheless,
considering the small and finite number of $n$ impurities in our samples with presumably $J = 1/2$
contributing to the magnetic entropy, we have $S_{mag} = 0.021$ J/mol K = $nR\ln(2)$, and therefore $n
\sim 0.4 \%$. Overall, our magnetic and thermodynamic characterization is consistent with a
non-magnetic ground state of Sr$_2$YIrO$_6$ with the presence of a small amount of correlated $J =
1/2$ paramagnetic centers. From our results, it is clear that the low temperature specific heat
anomalies are not related with the onset of long-range magnetic order.

\section{\label{sec:conc} Conclusions}

Following a recent matter of debate as to the evolution of magnetism in monoclinic Sr$_2$YIrO$_6$, we
have grown single crystals of Sr$_2$YIrO$_6$ using anhydrous SrCl$_2$ flux. Our single crystal XRD
results shows for the first time the presence of a $\sqrt{2}a \times \sqrt{2}a \times 1c$ supercell,
where $a$, $b$ and $c$ are the unit cell dimensions of the monoclinic subcell, highlighting the cubic
supercell of this compound.

The magnetic susceptibility revealed a predominantly paramagnetic
behavior in the temperature range 0.4 K $<$ T $\leq$ 300 K with an
effective magnetic moment $\mu_{eff} = 0.21 \mu_B$/Ir and
non-negligible but small spin correlations. From the Brillouin fit
for the isothermal magnetization curve at T = 0.48 K a low
percentage of $J = 1/2$ impurities were identified as the source
for the observed paramagnetism. No long-range magnetic ordering
was observed down to 430~mK.

A thorough study of the specific heat was carried out, in which a
broad anomaly in the low temperature region T $\leq$ 5 K was
identified as a Schottky anomaly caused by paramagnetic impurities
present in the sample, which are of the same order of magnitude
($n \sim 0.4 - 0.7$ \%) as obtained from magnetization
measurements. Our results are in strong contrast with the reported
ones by Cao et al. \cite{Cao}, but in full agreement with recent
reports on polycrystalline samples \cite{Ranjbar, Phelan},
pointing towards a non-magnetic ground state with the presence of
a small amount of correlated $J = 1/2$ impurities.


\section*{Acknowledgments}

The authors would like to thank M. Vojta, D. Efremov, J. van den Brink and J. Wosnitza for fruitful
discussions. This work has been supported by the Deutsche Forschungsgemeinschaft DFG under SFB 1143
and the Emmy-Noether program (Grant No. WU595/3-3).


\begin{thebibliography}{10}
\bibitem{5T_note} Note that the tiny humps in the 5 T curve are due to technical artifacts which are arising from the combination of the large number of data points and
negligible changes in the susceptibiity in the shown temperature regime.
\bibitem{g_note} A Brillouin function with $J = 3/2$ also allows to fit the data. However, this fit leads to non-consistent, too low g values.
\end{thebibliography}
\end{document}